A novel plate-type phononic crystal for efficient vibration and noise attenuation performance

Hao Zhu[1], Kai Zhao[1,2]*

1. Jiangsu Key Laboratory of Advanced Food Manufacturing Equipment and Technology, Wuxi 214122, China

2. School of Mechanical Engineering, Jiangnan University, Wuxi 214122, China

Mitigating low-frequency vibration or noise is of vital importance to both human health and mechanical engineering. Two-dimensional phononic crystal (PC) structures were proposed by attaching rubber and metallic cylinders on one or both sides of a thin plate to attenuate low-frequency vibration via the local resonance mechanism. The finite element method was employed to evaluate the band structure and associated vibration modes of the proposed PC structures. It was found that the bandgap of the single-sided PC structure is narrower than that of the double-sided configuration. The formation mechanism of the flexural wave bandgap was analyzed based on the vibration modes. In particular, the influence of structural and material parameters on the band structure was systematically investigated. Finally, the accuracy of the calculated band structure and the effectiveness of the vibration and noise reduction performance were verified through simulations of vibration transmission characteristics and sound insulation curves. The results indicate that the proposed PC generates a relatively wide flexural wave bandgap in the low-frequency range, whose width and position can be flexibly tuned by adjusting structural and material parameters. These findings provide a novel approach for controlling low-frequency vibration and noise.

**Keywords**: Local resonance; Phononic crystal; Band structure; Vibration modes; Transmission characteristics

* Corresponding author.
*E-mail address*: kai.zhao@jiangnan.edu.cn (K.Z.)

# 1. Introduction

Noise and vibration are commonly encountered in daily life and industrial environment. The low-frequency components, characterized by strong penetration and slow attenuation, not only pose risks to human physiological systems but also disrupt the normal functioning of precision instruments and equipment[1-3]. Consequently, considerable efforts have been devoted to address this issue. Traditional vibration and noise mitigation structures are limited by the mass law, which states that their damping performance is positively correlated with structural density[4, 5]. Effective control of low-frequency noise thus necessitates the use of larger structures, which are often impractical in real-world applications. Therefore, it is essential to develop novel materials that can overcome the limitations of conventional vibration and noise control methods while effectively addressing low-frequency noise attenuation.

Acoustic metamaterials[6-8], also referred to as phononic crystals, are artificial structures consisting of periodically arranged units that are specifically designed to manipulate the propagation of elastic (acoustic) waves within a medium. Specifically, the bandgaps formed between dispersion curves in phononic crystals can effectively suppress vibrations and insulate sound. Early studies of phononic crystals were mainly based on the Bragg scattering mechanism[9-11], which has a wide frequency range and high attenuation capability for the Bragg scattering bandgap, but their engineering applications are limited due to the requirement of wavelengths comparable to the crystal lattice constants, thus they are usually applied at high frequencies. To address the limitations of Bragg scattering in low-frequency vibration and noise attenuation, local resonators have been incorporated into phononic crystals to induce sub-wavelength bandgaps. In a pioneering study, Liu et al.[12] designed resonant units by coating high-density lead spheres with viscoelastic silicone rubber, which were subsequently embedded in an elastic epoxy resin matrix. Since then, locally resonant phononic crystals (LRPCs) have offered innovative solutions for low-frequency vibration and noise mitigation. Li et al.[13] and Wu et al.[14] employed phononic crystals for noise control in automobile interiors, demonstrating that LRPCs can effectively reduce interior noise. Ruan et al.[15], Qian et al.[16], and Zhang et al[17] found that phononic crystals could significantly attenuate vibration and noise in ship structures. Additionally, researches in rail transportation by Qu et al.[18] and Zhao et al.[19] demonstrated that floating plate rail systems incorporating phononic crystals outperform traditional steel spring systems.

Since the bandgap characteristics of LRPCs are closely dependent on their structural configurations, the design of their geometric shapes is essential to achieve the optimal performance. Xiao et al.[20] developed a cookie-shaped phononic crystal whose bandgap width could be adjusted by altering geometric parameters. Yang et al.[21] demonstrated that a ring-slotted spiral phononic crystal structure generates a bandgap ranging from 23.54 to 102.87 Hz. Li et al.[22] designed a radial plate-type locally resonant phononic crystal. By simultaneously tuning the material composition and spatial distribution of the wrapping layer, they achieved bandgap widths spanning three orders of magnitude. Gupta et al.[23] proposed an hourglass-shaped phononic crystal with a parcel layer that facilitates the formation of a varying stiffness profile, thereby enabling the structure to generate an ultrabroad bandgap with enhanced vibration attenuation.

Although extensive research has demonstrated acoustic metamaterials for controlling low-frequency structural vibrations and mitigating underwater[24-26] and air noise[27-30] in standard engineering systems, their structural design faces significant challenges when lightweight requirements are imposed by specialized applications. How to design a phononic crystal structure that simultaneously satisfies the low-frequency, broadband, and lightweight as a whole is still a shortcoming of current researches. To achieve low-frequency vibration damping and noise reduction under lightweight design constraints, a novel type of LRPC structure is proposed in this study. The vibration damping and noise mitigation performance of the proposed structure is comprehensively investigated using the finite element method (FEM) simulations. The structure of this paper is organized as follows: **Section 2** presents the structural configuration and material properties of the phononic crystal, and introduces the FEM procedures for calculating the band structure. **Section 3** analyzes the mechanisms of bandgap formation and examines the effects of critical structural and material parameters on the bandgap in details. **Section 4** investigates, through numerical simulations, the vibration transmission characteristics and sound insulation performance of the periodic phononic crystal plate.

## 2. Phononic crystal modeling and computational methods

### 2.1. Structural design

In this study, two-dimensional LRPC structures were constructed by attaching cladding layers and metallic scatterers to one or both sides of a substrate plate. The substrate plate was square, while the cladding layers and scatterers were modeled as solid cylinders. The geometric parameters of the phononic crystal are listed in **Table 1**, while the material properties are presented in **Table 2**.

Table 1. Geometric settings of the unit cell shown in **Fig. 1**.

| Parameters | Size (mm) |
|---|---|
| $a$ | 20 |
| $b$ | 18 |
| $c$ | 12 |
| $d$ | 3 |
| $h_1$ | 3 |
| $h_2$ | 3 |
| $h_3$ | 10 |
| $r_1$ | 6 |
| $r_2$ | 6 |
| $r_3$ | 2 |

Table 2. Material parameters of the LRPC unit cell.

| Component | Material | Density (kg/m$^3$) | Young's modules (GPa) | Poisson's ratio |
|---|---|---|---|---|
| Substrate plate | Epoxy resin | 1180 | 4.35 | 0.37 |
| Cladding layer | Rubber | 1300 | 1.175×10$^{-4}$ | 0.47 |
| Scatterer | Steel | 8750 | 200 | 0.33 |

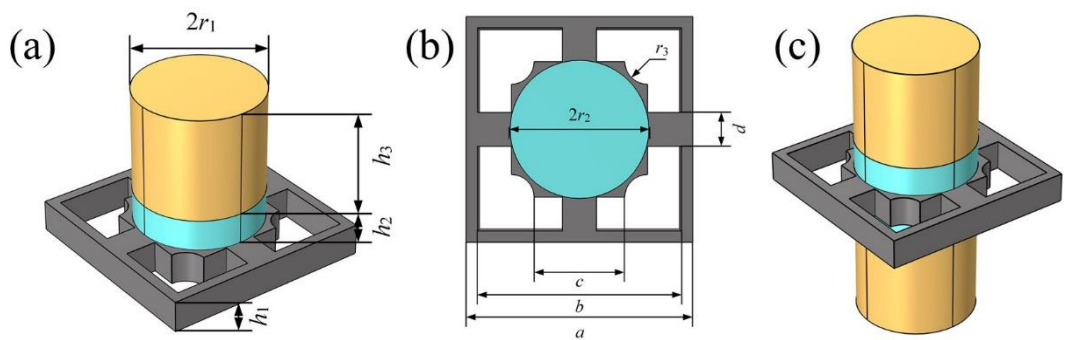

**Fig. 1**. Single-cell structure of the phonon crystal, (a) single-sided phononic crystal structure, (b) substrate plate and cladding layer, (c) double-sided phononic crystal structure.

*2.2. Calculation method*

The propogation of an elastic wave in a linearly elastic, isotropic, infinite-volume, undamped and homogeneous medium is given by

$$\mu\nabla^2 u(\mathbf{r}) + (\lambda + \mu)\nabla(\nabla \cdot u(\mathbf{r})) = -\rho\omega^2 u(\mathbf{r}) \tag{1}$$

where, $u(r)$ is the displacement vector, $\nabla$ is the Hamiltonian, $\rho$ is the material density, while $\lambda$ and $\mu$ are Lame constants of the medium. Due to the periodicity and symmetry of phononic crystal, only primitive cells should be considered when calculating the energy band structure. According to the Bloch's theorem, the cell-node displacement field function is:

$$u(\mathbf{r}) = u_k(\mathbf{r})e^{i(\mathbf{k}\cdot\mathbf{r})} \tag{2}$$

where, $u_k(r)$ is periodic within the unit cell, with $k$ is the first irreducible Brillouin zone wave vector. For the FEM calculation, the overall model is discretized first. Then stiffness and mass matrices are assembled for the unit cell nodes. The eigenequations of the primitive cell can thus be derived as:

$$(K - \omega^2 M)U = 0 \tag{3}$$

where, $K$ is the cell stiffness matrix, $M$ is the mass matrix, and $U$ is the eigenvector. According to the Bloch's theorem, the outer boundary of the primitive cell adopts the Bloch-Floquet periodic boundary condition, *i.e.*:

$$u(r+a) = e^{i(k-r)}u(r) \tag{4}$$

where, $r$ is the position vector of the boundary node, and $a$ is the lattice constant.

For each given Bloch wave vector $k$, the corresponding eigenfrequencies can be obtained by solving eigenequations (Eq.-(3)). Since the phononic crystal has special point group symmetry properties, the wave vector $k$ needs to be confined on the boundary of the irreducible Brillouin zone (**Fig. 2**) to determine the energy band structure of the phononic crystal.

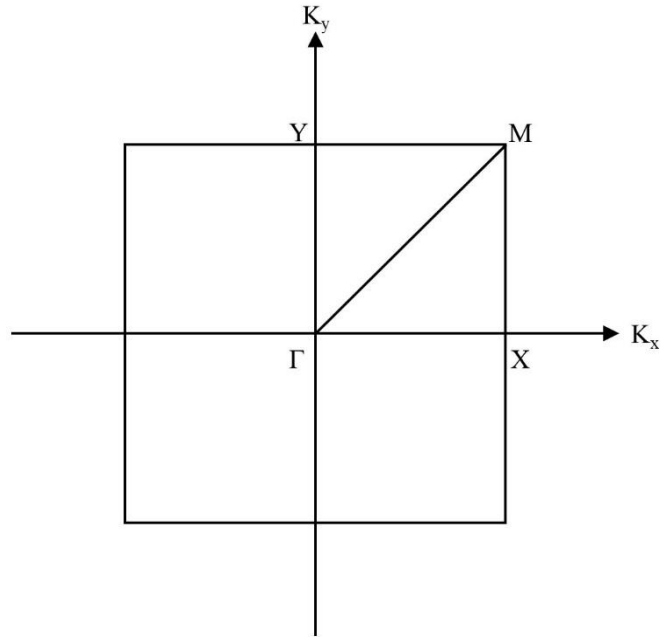

**Fig. 2**. Irreducible Brillouin zone.

## 3. Bandgap characterization of phononic crystals

### 3.1. Band structure

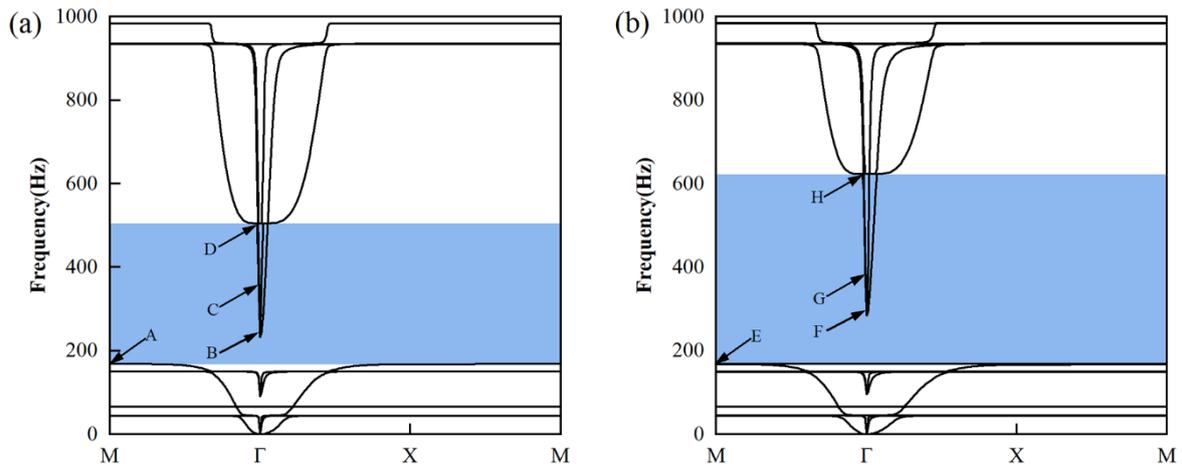

**Fig. 3**. Band structure of the, (a) single-sided phononic crystal, and (b) double-sided phononic crystal.

According to the elastodynamic theory for solid plates, guided wave propagation manifests as two fundamental modes: horizontal shear waves and Lamb waves. Lamb waves further bifurcate into symmetric and antisymmetric modes, with the fundamental antisymmetric mode conventionally termed flexural waves in thin-plate regimes. In thin plates, the motion directions of horizontal shear waves and symmetric Lamb waves are parallel to the plate surface, whereas the motion of flexural waves is perpendicular to it. The vibration of the plate primarily consists of longitudinal (parallel to the surface) and flexural components

(perpendicular to the surface), leading to the formation of both longitudinal and flexural bandgaps in plate-type phononic crystals. The present study mainly focuses on suppressing flexural waves and, consequently, analyzes the corresponding bandgaps. Given that horizontal shear and symmetric Lamb waves typically exhibit linear propagation behavior at low frequencies, the analysis of the band structure and vibration modes of the phononic crystal is used to identify the flexural bandgaps. As shown in **Fig. 3**, for the single-sided phononic crystal, the flexural wave bandgap ranges from 168.3 Hz to 505.02 Hz, yielding a bandgap width of 336.72 Hz. In the case of the double-sided phononic crystal, the bandgap spans from 168.9 Hz to 622.58 Hz, with a width of 453.68 Hz.

To investigate the mechanism underlying the formation of the flexural wave bandgap, eight points (A–H) within the frequency range of the bandgap were selected for modal analysis based on the band structure diagram (see **Fig. 3**, where red arrows indicate displacement directions). Points A–D correspond to vibration modes of the single-sided phononic crystal, while points E–H represent those of the double-sided structure. The mode geometries and corresponding frequencies of points A–H are presented in **Fig. 4**. Modes A and E primarily demonstrate vibration of the scatterer along the $Z$ direction, stretching of the cladding layer, and effectively immobilized substrate behaviors. At this stage, the translational motion of the scatterer exerts a reaction force along the $Z$ direction on the substrate, thereby suppressing flexural wave and inhibiting its propagation, leading to the formation of the bandgap. Modes B, C, F, and G indicate that the scatterer remains nearly stationary, while the substrate exhibits translational motion in the *xOy* plane. In these modes, longitudinal vibrations in the *xOy* plane propagate through the substrate; however, vibrations along the $Z$ direction are effectively suppressed, thus the flexural wave cannot propagate within the bandgap region shown in **Fig. 3**. Modes D and H are characterized by a nearly stationary scatterer, with $Z$-direction oscillations in the substrate, and compression of the cladding layer. In this case, no reaction force is generated by the scatterer along the $Z$ direction, thereby allowing the flexural wave to propagate through the substrate.

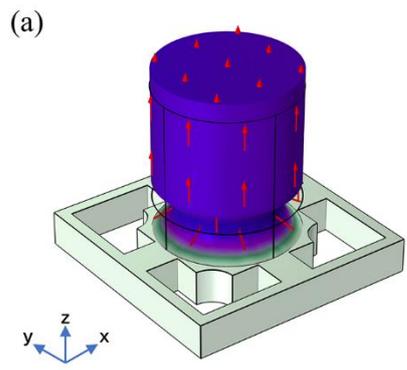
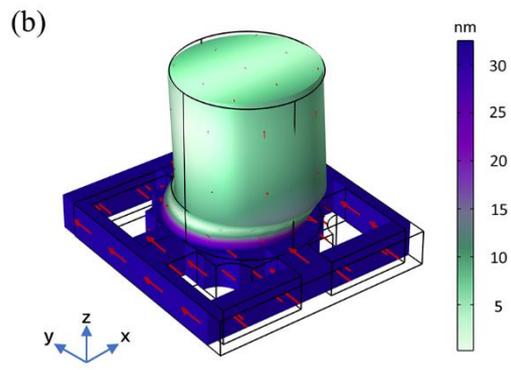
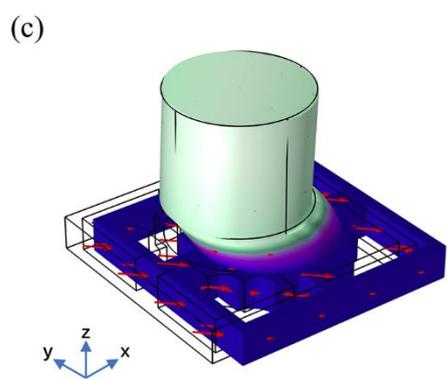
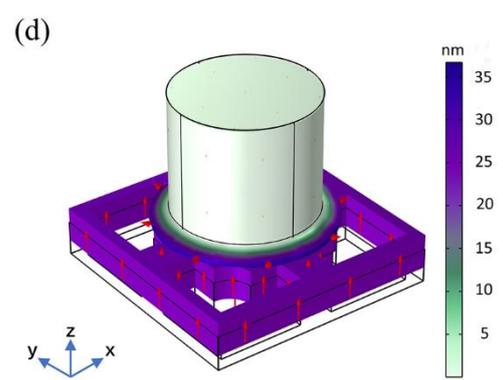
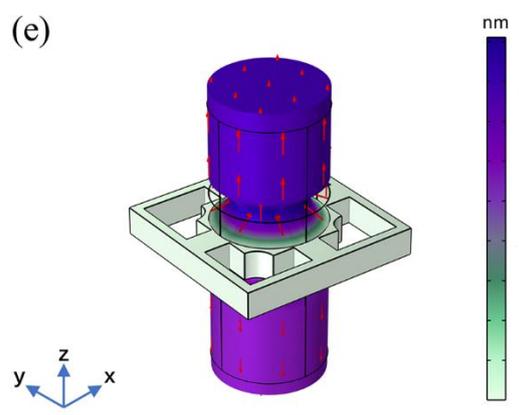
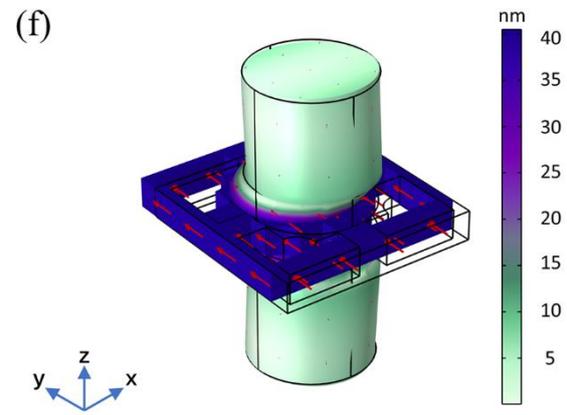
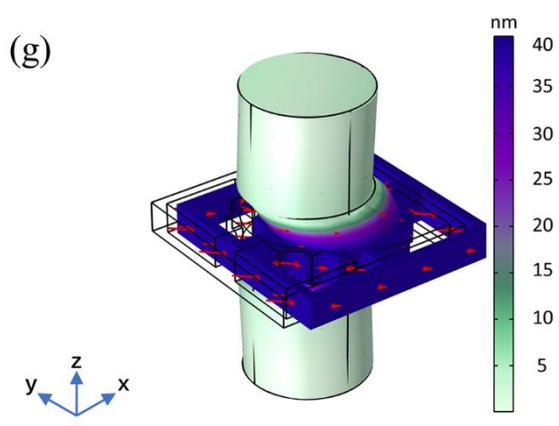
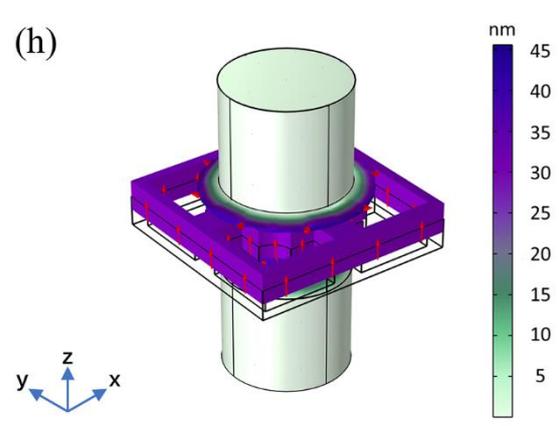

**Fig. 4**. Plot of A-H modal vibration patterns, (a) point A (168.3Hz), (b) point B (245.97Hz), (c) point C (359.82Hz), (d) point D (505.02Hz), (e) point E (168.9Hz), (f) point F (295.34Hz), (g) point G (383.05Hz), (h) point H (622.58Hz).

### 3.2. Factors Influencing Bandgaps

Knowing the bandgap properties of phononic crystals is crucial for effective vibration and noise reduction. Therefore, a control variable experimentation is designed to investigate the effects of structural and material parameters on the bandgap of both single-sided and double-sided phononic crystals.

Structural parameters (scatterer radius/height, substrate/cladding thickness) are examined, with their impact on the band structure demonstrated in **Fig. 5**.

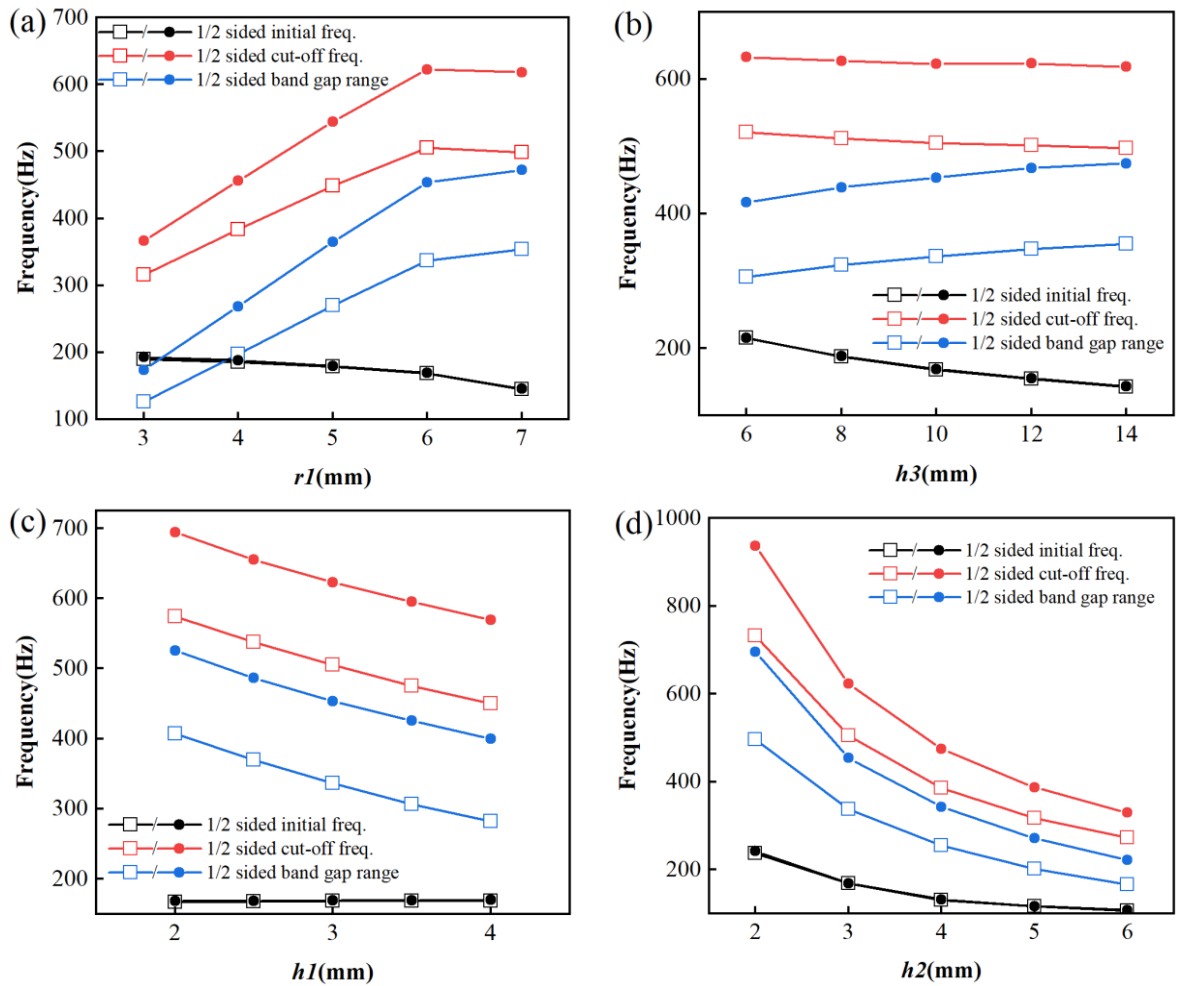

**Fig. 5**. Influence of structural parameters, (a) scatterer radius $r_1$, (b) scatterer height $h_3$, (c) substrate thickness $h_1$, and (d) cladding layer thickness $h_2$ on band gap.

As shown in **Fig. 5**(a), increasing the scatterer radius leads to a decrease in the initial frequency of the bandgap, while the cutoff frequency increases significantly, resulting in a substantial change in the bandgap width. As illustrated in **Fig. 5**(b), an increase in the scatterer height causes both the initial and cutoff frequencies of the bandgap to decrease slightly, indicating a limited impact on the overall bandgap width. **Fig. 5**(c) shows that, increasing the substrate thickness causes a slight rise in the initial frequency and a rapid decrease in the cutoff frequency, leading to a significant reduction in the bandgap width. **Fig. 5**(d) illustrates that, both the initial and cutoff frequencies of the bandgap exhibit a rapid downward trend with increasing cladding layer thickness, resulting in a noticeable narrowing of the bandgap.

To examine the influence of material properties on the flexural wave bandgap of the phononic crystal, we further analyze the effects of the scatterer, cladding layer, and substrate materials in sequence. **Table**s **3**, **4**, and **5** enumerate the material parameters of the scatterer, cladding layer, and substrate, respectively, while the simulation results are presented in **Fig. 6**.

Table 3 Material parameters of the scatterer.

| Name | Density (kg·m$^{-3}$) | Young's modulus (GPa) | Poisson's ratio |
|---|---|---|---|
| Steel | 7850 | 200.0 | 0.33 |
| Copper | 8960 | 127.1 | 0.35 |
| Lead | 11600 | 40.8 | 0.37 |
| Tungsten | 19100 | 354.1 | 0.35 |

Table 4 Material parameters of the cladding layer.

| Name | Young's modulus (GPa) |
|---|---|
| Rubber1 | $1.175 \times 10^{-4}$ |
| Rubber2 | $1.37 \times 10^{-4}$ |
| Rubber3 | $7.7 \times 10^{-4}$ |
| Rubber4 | $10 \times 10^{-4}$ |

Table 5 Material parameters of the substrate plate.

| Name | Density (kg·m$^{-3}$) | Young's modulus (GPa) | Poisson's ratio |
|---|---|---|---|
| Epoxy resin | 1180 | 4.35 | 0.37 |
| Plastic | 1190 | 2.2 | 0.38 |
| Aluminum | 2700 | 70 | 0.35 |
| Steel | 7850 | 200 | 0.33 |

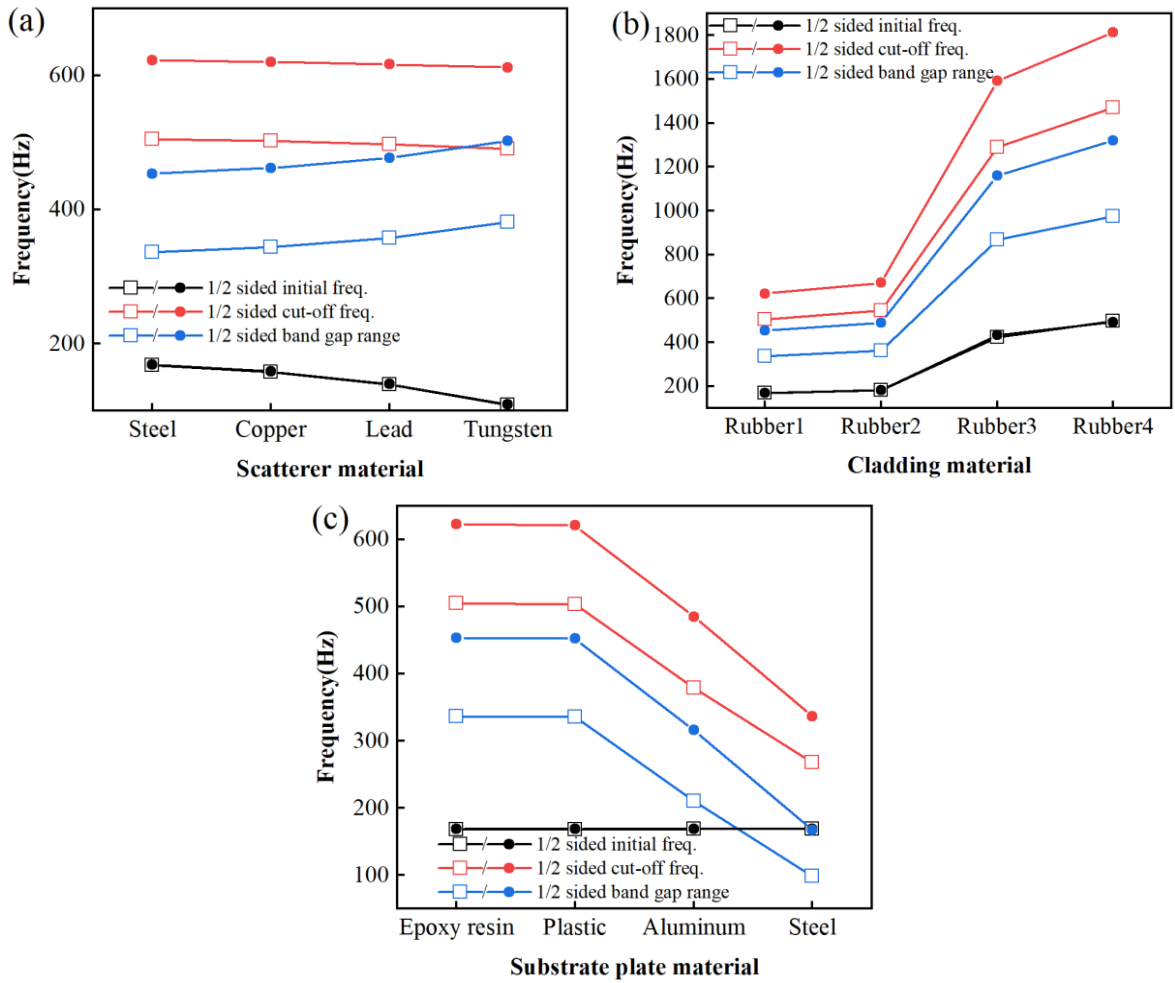

Fig. 6. Effect of material parameters, (a) the scatterer, (b) the cladding layer, (c) the substrate plate on band gap.

As shown in **Fig. 6**(a), when the scatterer materials are sequentially selected as steel, copper, lead, and tungsten (corresponding to increasing material density), the cutoff frequency of the bandgap remains nearly constant, whereas the initial frequency decreases sharply, resulting in a gradual increase in the bandgap width. **Fig. 6**(b) illustrates that, an increase in the Young's modulus of the cladding material leads to upward shifts in both the initial and cutoff frequencies, thereby expanding the bandgap and shifting it toward the higher-frequency region. **Fig. 6**(c) shows that, when the substrate materials are sequentially varied from epoxy resin to plastic, aluminum, and steel, the initial frequency of the bandgap remains nearly unchanged, while the cutoff frequency decreases significantly.

The preceding analysis indicates that both structural and material parameters significantly affect the bandgap characteristics of the proposed plate-type phononic crystals. Based on the observed trends in the band structure, the bandgap can be effectively tuned through appropriate parameters' selection to meet requirements of specific applications.

## 4. Transmission properties

*4.1. Vibration transmission calculations*

Based on the band structure of phononic crystals, elastic waves within the bandgap frequency range are attenuated due to local resonance effects. However, in practical applications, the effectiveness and extent of vibration isolation produced by phononic crystals require further investigation. Therefore, the vibration transmission characteristics of the phononic crystal plate is further evaluated using FEM calculations. The computational model, illustrated in **Fig. 7**, applies a *Z*-direction displacement excitation on the left-side edge (indicated by black arrows), while the displacement response is measured on the right-side edge. The flexural wave transmission loss (TL) can be calculated using the following equation:

$$TL = 20\lg \frac{w_{out}}{w_{in}} \tag{5}$$

where, $w_{in}$ is the displacement excitation, $w_{out}$ is the displacement response.

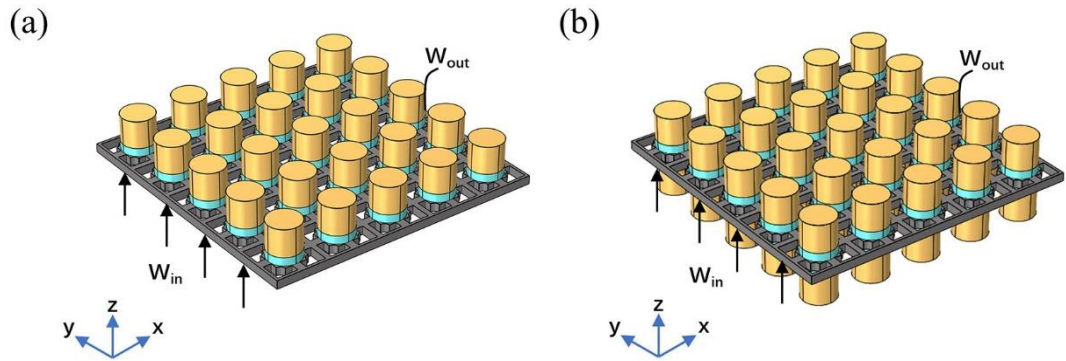

**Fig. 7**. Finite-size periodic structures of: (a) single-sided phononic crystal, (b) double-sided phononic crystal.

To investigate the vibrational transmission behavior of the phononic crystal plate, transmission curves were computed for systems with different numbers of unit cells. As shown in **Fig. 8**(a), a sharp attenuation is observed between 172 Hz and 538 Hz, closely matching the flexural wave bandgap range (168.3-505.02 Hz) of the single-sided phononic crystal. Similarly, **Fig. 8**(b) demonstrates that the flexural wave attenuation range in the transmission spectrum agrees well with the bandgap of the double-sided phononic crystal. Moreover, the attenuation of the flexural wave increases progressively with the number of unit cells. The transmission spectrum not only validates the accuracy of the calculated band structure but also confirms the phononic crystal's capability to effectively suppress flexural wave propagation within the bandgap.

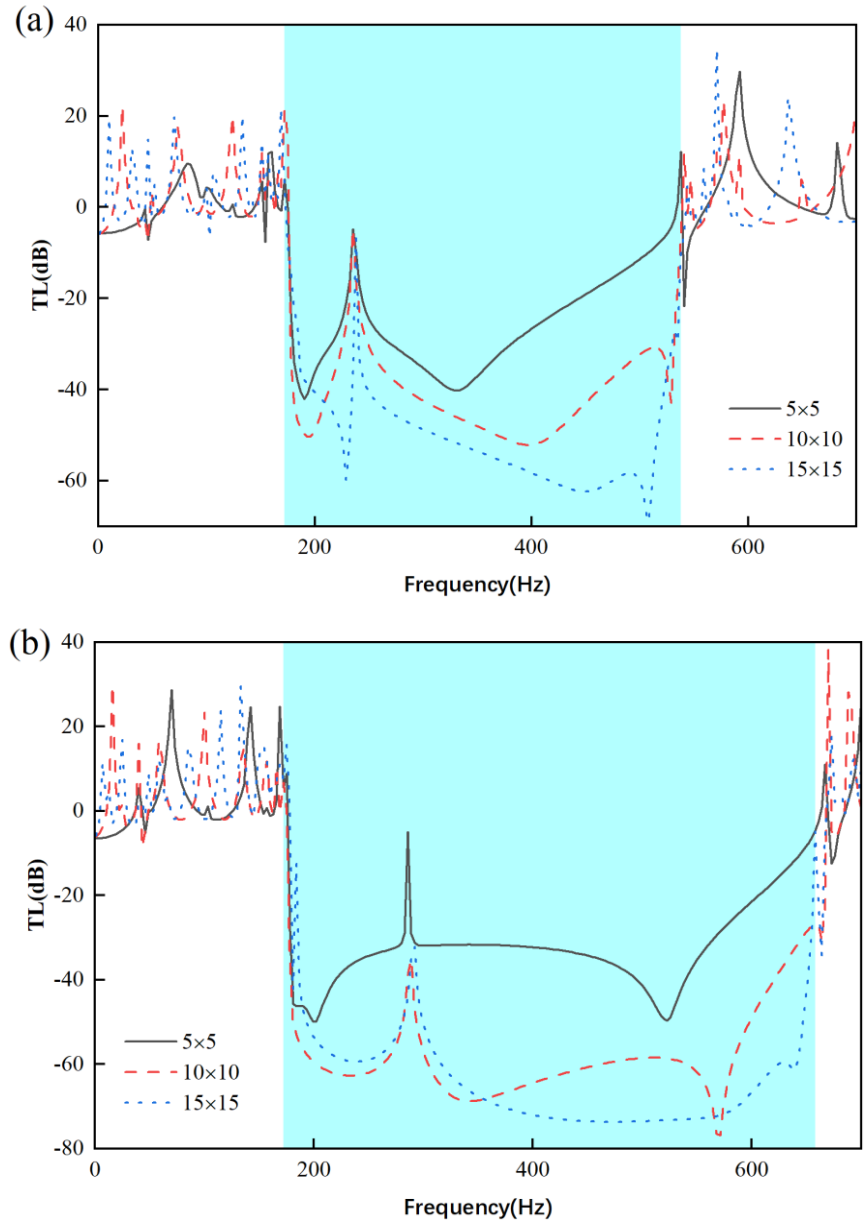

**Fig. 8**. Calculated transmission characteristic curves of, (a) single-sided phononic crystal, (b) double-sided phononic crystal.

To provide a more intuitive demonstration of the vibration damping performance of the phononic crystal plate, a 5×5 array structure is selected as the representative example without loss of generality. Vibration displacement contour diagrams corresponding to frequencies within and outside the bandgap range for both single-sided and double-sided phononic crystals are presented in **Fig. 9** and **Fig. 10**, respectively. As clearly observed from the vibration displacement diagrams, when the excitation signal frequency lies within the bandgap range, the vibration energy is effectively blocked by the phononic crystal plate, preventing the propagation of the flexural wave and resulting in the significant vibration isolation. In contrast, when the excitation frequency falls outside the bandgap, the phononic crystal plate does not impede the vibration energy, allowing the flexural wave to propagate through the plate.

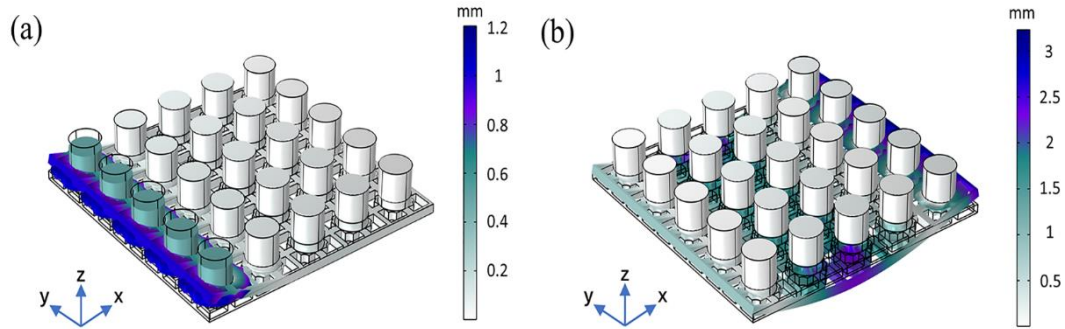

**Fig. 9**. Vibration displacement contour diagrams of the single-sided phononic crystal plate: (a) 260 Hz, (b) 600 Hz.

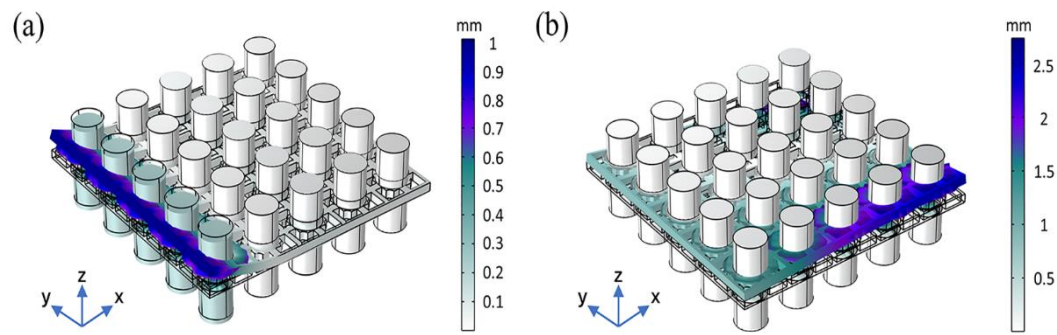

**Fig. 10**. Vibration displacement contour diagrams of the double-sided phononic crystal plate, (a) 273 Hz, (b) 665 Hz.

Phononic crystals are commonly applied to the surfaces of structures requiring vibration and noise suppression to achieve efficient control. However, real-world engineering constraints introduce uncontrolled variables. For instance, the existence of surface geometric imperfections (e.g., roughness, warpage), could compromise interface flatness and disrupt phononic crystal unit alignment, thus hinders the proper arrangement of phononic crystal units. Additionally, prolonged services might lead to the detachment or loss of metallic scatterers. The absence of certain unit cells and the loss of metallic scatterers can significantly degrade the overall vibration damping performance of the phononic crystal array. To evaluate this effect, the vibration transmission characteristics of single-sided phononic crystal plates with different numbers of missing unit cells and scatterers are analyzed. The defected models are illustrated in **Fig. 11**, and the corresponding vibration attenuation results are presented in **Fig. 12** and **Fig.**

**13**, respectively.

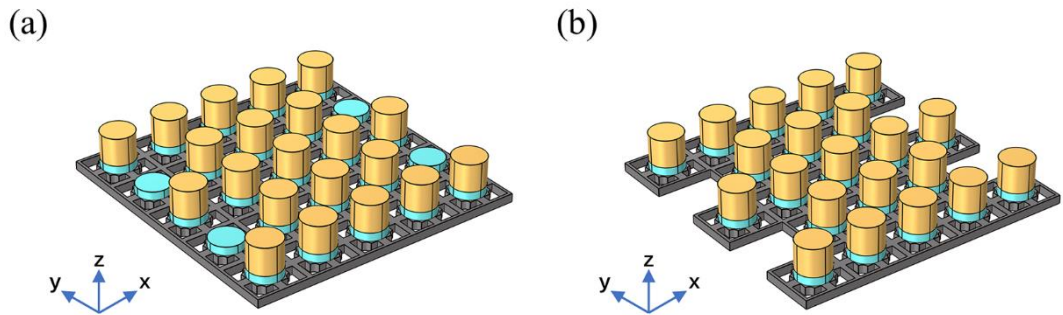

**Fig. 11**. Models of defected phonon crystal plates, (a) with unit cells missed, (b) with scatters missed.

As shown in **Fig. 12**, reducing the number of unit cells degrades flexural wave attenuation within the bandgap, directly compromising the efficacy of vibration isolation. To maximize performance, dense phononic crystal arrays should be deployed within feasible geometric constraints. As shown in **Fig. 13**, the absence of metallic scatterers not only decreases the magnitude of flexural wave attenuation but also narrows the effective bandwidth. Hence, maintaining the integrity of the phononic crystal is essential under service conditions.

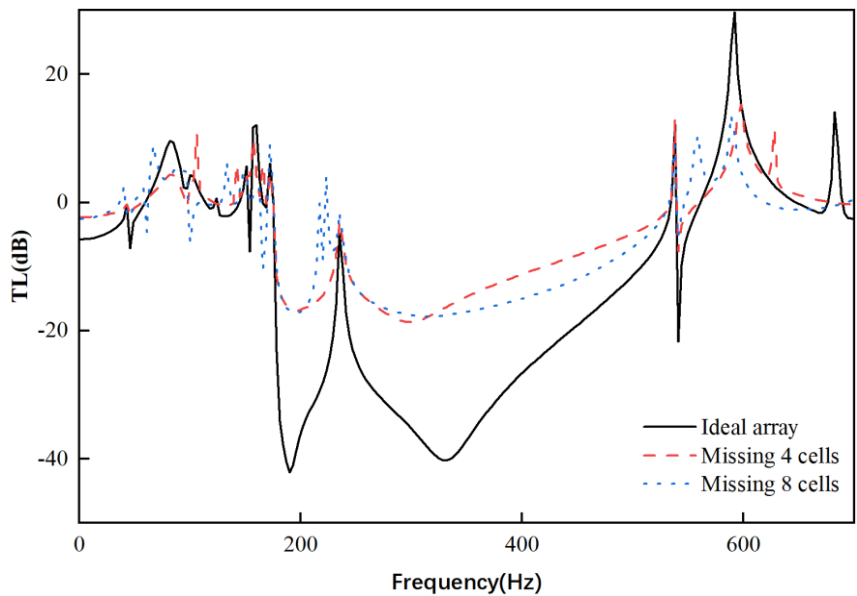

**Fig. 12**. Comparison of vibrational attenuation of phononic crystal plate with different numbers of unit cells missed.

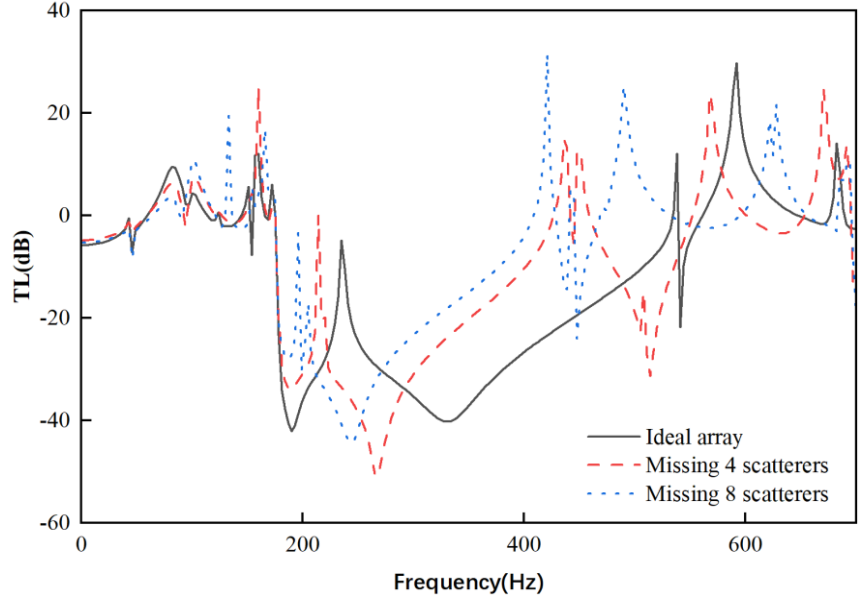

Fig. 13. Comparison of vibrational attenuation of phononic crystal plate with different numbers of metallic scatters missed.

*4.2. STL curve calculation*

To evaluate the sound insulation performance of the proposed phononic crystal structure, sound transmission loss (STL) curves are calculated. The computational model is illustrated in **Fig. 14**. The central region of the model represents the sound-insulating structure, while air layers and perfectly matched layers (PMLs) are placed on both sides to simulate open boundary conditions. Periodic boundary conditions are applied along the lateral sides of the model to simulate an infinite array configuration. A background acoustic pressure field is imposed on the upper air layer to serve as the excitation source. To determine the STL, acoustic pressure monitoring planes are established at the air-solid interfaces on both sides. The STL value is computed via the ratio of incident to transmitted sound power as follows:

$$STL = 10\lg\left(\frac{\int \frac{p_0^2}{2\rho c} dA}{\int \frac{|p|^2}{2\rho c} dA}\right) \quad (6)$$

where, $p_0$ is the incident sound pressure amplitude, $p$ is the transmitted sound pressure, $\rho$ is the air density, $c$ is the sound velocity.

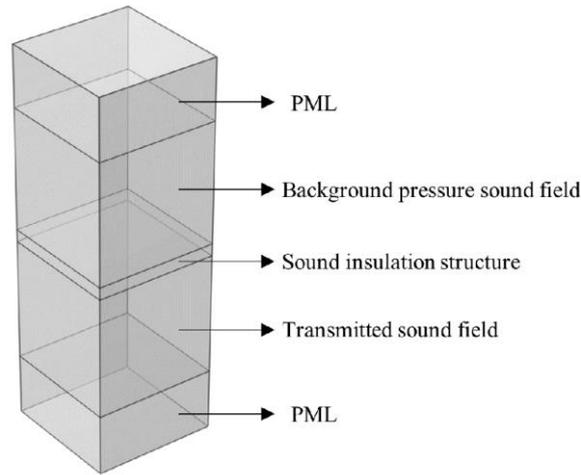

**Fig. 14**. The STL calculation model.

The STL curves for the phononic crystal are shown in **Fig. 15**, with that of the epoxy resin plate is also given for comparison.

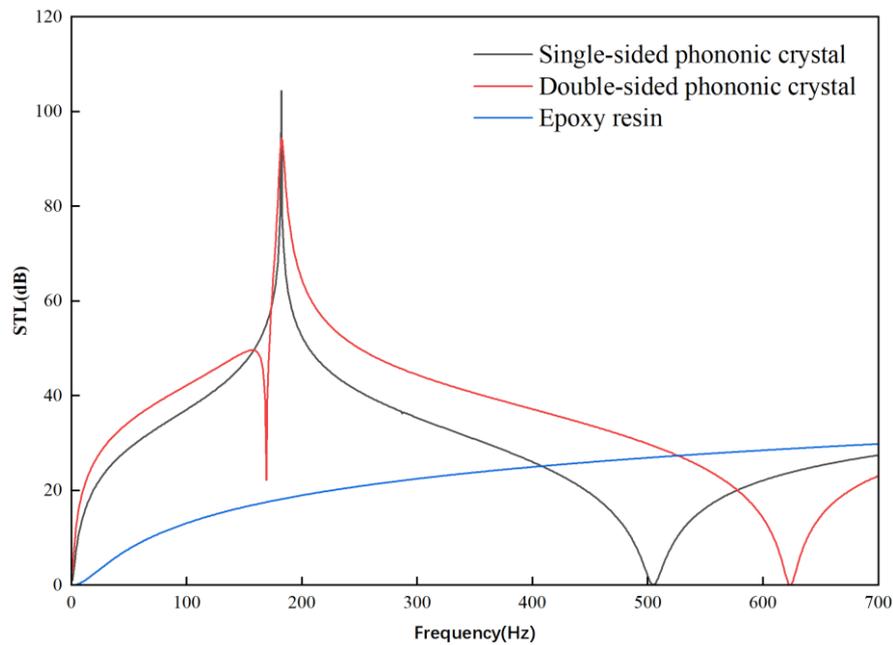

**Fig. 15**. STL curves of phononic crystals and epoxy resin plate.

As shown in **Fig. 15**, the average STL within the bandgap of the single-sided phononic crystal reaches 32 dB, with a maximum value of 104 dB. For the double-sided phononic crystal, the average STL is 38 dB, with a maximum of 94 dB. By contrast, the epoxy resin plate exhibits an average STL of only 21 dB and a maximum of 29 dB. These results indicate that the phononic crystal structures exhibit superior sound insulation performance over a wide frequency range compared to the epoxy resin substrate, effectively meeting the requirement of acoustic insulation within the target bandgap range.

To further elucidate the sound isolation mechanism, the displacement and sound pressure contours of the single-sided phononic crystal at the maximum and minimum points of the STL curve are shown in **Fig. 16**. At 182 Hz, the motion of the unit cell is primarily contributed by the displacement of the metallic scatter along the *Z*-direction, while the substrate remains stationary, consistent with the eigenmode at the initial frequency of the flexural wave bandgap (**Fig. 4**(a)). Furthermore, the sound pressure distribution reveals a significant reduction of pressure amplitude between the source-side and the transmitted-side regions. At 505 Hz, the metallic scatterer is nearly stationary, while the substrate moves along the *Z*-direction, consistent with the eigenmode at the cutoff frequency of the flexural wave bandgap (**Fig. 4**(d)). The sound pressure contour reveals that the pressure remains nearly constant in both the incident and transmitted fields at this frequency. Based on the above analysis, it can be concluded that the phononic crystal exhibits high sound insulation efficiency within the flexural wave bandgap.

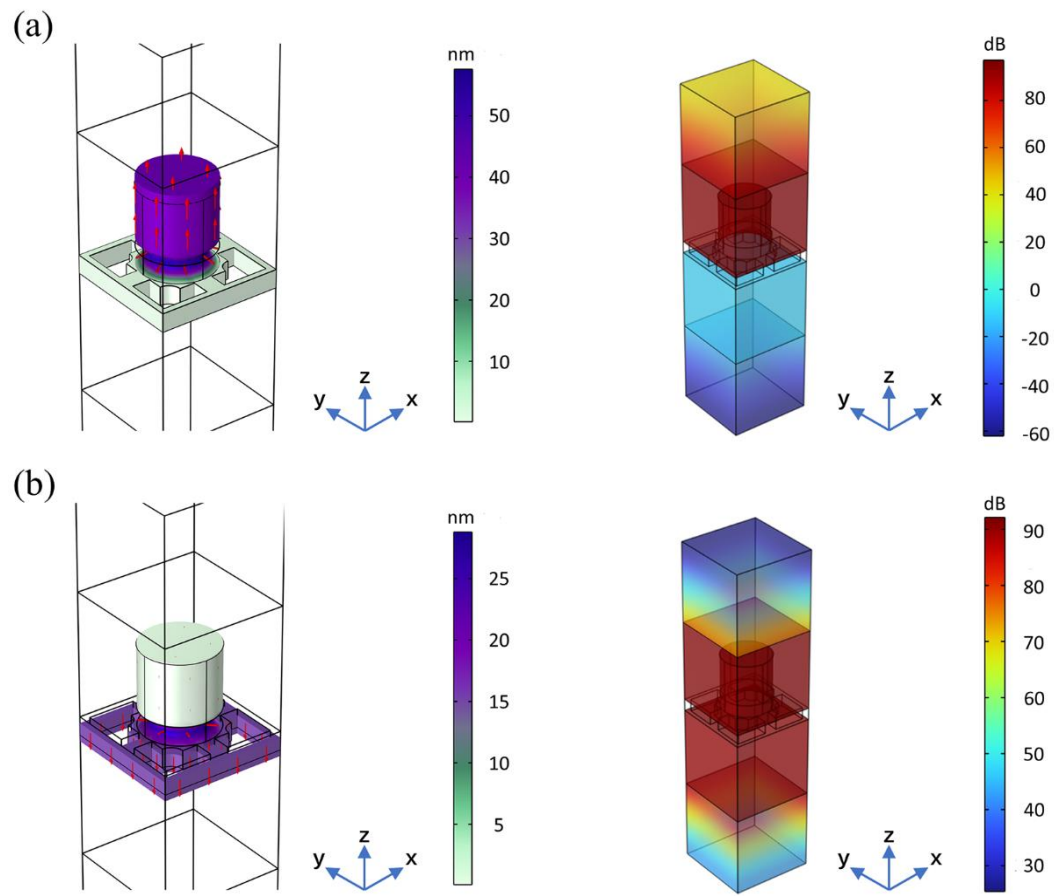

**Fig. 16**. The displacement and sound pressure contours at the minimum and maximum points of STL curve of the single-sided phononic crystal, (a) 182 Hz, (b) 505 Hz.

## 5. Conclusions

In summary, the energy band structures of both single-sided and double-sided phononic crystal configurations are investigated using FEM analysis. It is demonstrated that the proposed structure can generate a wide flexural wave bandgap at low frequencies, with its center frequency and bandwidth being tunable by varying structural and material parameters. The effectiveness of the proposed structure in suppressing vibration and noise is verified through analyses of vibration transmission and STL curves. The findings of this study might offer valuable insights for further applications of acoustic metamaterials in engineering practice.

**Acknowledgements**

Financial supports provided by the National Natural Science Foundation of China (Grant No. 12102145) and Natural Science Foundation of Jiangsu Province (Grant No. BK20210444) are acknowledged.